\documentstyle[aps,preprint,epsf,citesort]{revtex}
\draft

\newcommand{\beq}{\begin{equation}}
\newcommand{\eeq}{\end{equation}}
\newcommand{\bdis}{\begin{displaymath}}
\newcommand{\edis}{\end{displaymath}}
\newcommand{\bea}{\begin{eqnarray}}
\newcommand{\eea}{\end{eqnarray}}
\newcommand{\barr}{\begin{array}}
\newcommand{\earr}{\end{array}}

\begin{document}
\title
{Water-Like Anomalies for Core-Softened Models of Fluids:\\ 
Two-Dimensional Systems}

\author{A. Scala, M. Reza Sadr-Lahijany,
  N. Giovambattista,\\ S. V. Buldyrev, and H. E. Stanley}

\address{Center for Polymer Studies and Department of Physics \\
  Boston University, Boston, MA 02215 USA}

\date{12 October 2000}

\maketitle

\begin{abstract}

We use molecular dynamics simulations in two dimensions to investigate
the possibility that a core-softened potential can reproduce static and
dynamic anomalies found experimentally in liquid water: (i) the increase
in specific volume upon cooling, (ii) the increase in isothermal
compressibility upon cooling, and (iii) the increase in the diffusion
coefficient with pressure. We relate these anomalies to the shape of the
potential.  We obtain the phase diagram of the system and identify two
solid phases: a square crystal (high density phase), and a triangular
crystal (low density phase). We also discuss the relation between the
anomalies observed and the polymorphism of the solid. Finally, we
compare the phase diagram of our model system with experimental data,
noting especially the line of temperatures of maximum density, line of
pressures of maximum diffusion constant, and line of temperatures of
minimum isothermal compressibility.

\end{abstract}


\section{Introduction}
\label{secint}

Most liquids contract upon cooling and become more viscous with
pressure. This is not the case for the most important liquid on earth,
water. For at least 300 years it has been known that the specific volume of
water at ambient pressure starts to increase when cooled below
$T=4^\circ$C \cite{Newref1}. It is perhaps less known that the viscosity
of water decreases upon increasing pressure in a certain range of
temperatures \cite{WaterData}. Moreover, in a certain range of pressures
water exhibits an anomalous increase of compressibility, and hence of
density fluctuations, upon cooling. These anomalies are not restricted
to water but are also present in other liquids
\cite{Debenedetti-book,mishima,Yoshimura}.

\begin{figure}[htb]
\narrowtext \centerline{
\hbox {
  \vspace*{0.5cm}
  \epsfxsize=11.5cm
  \epsfbox{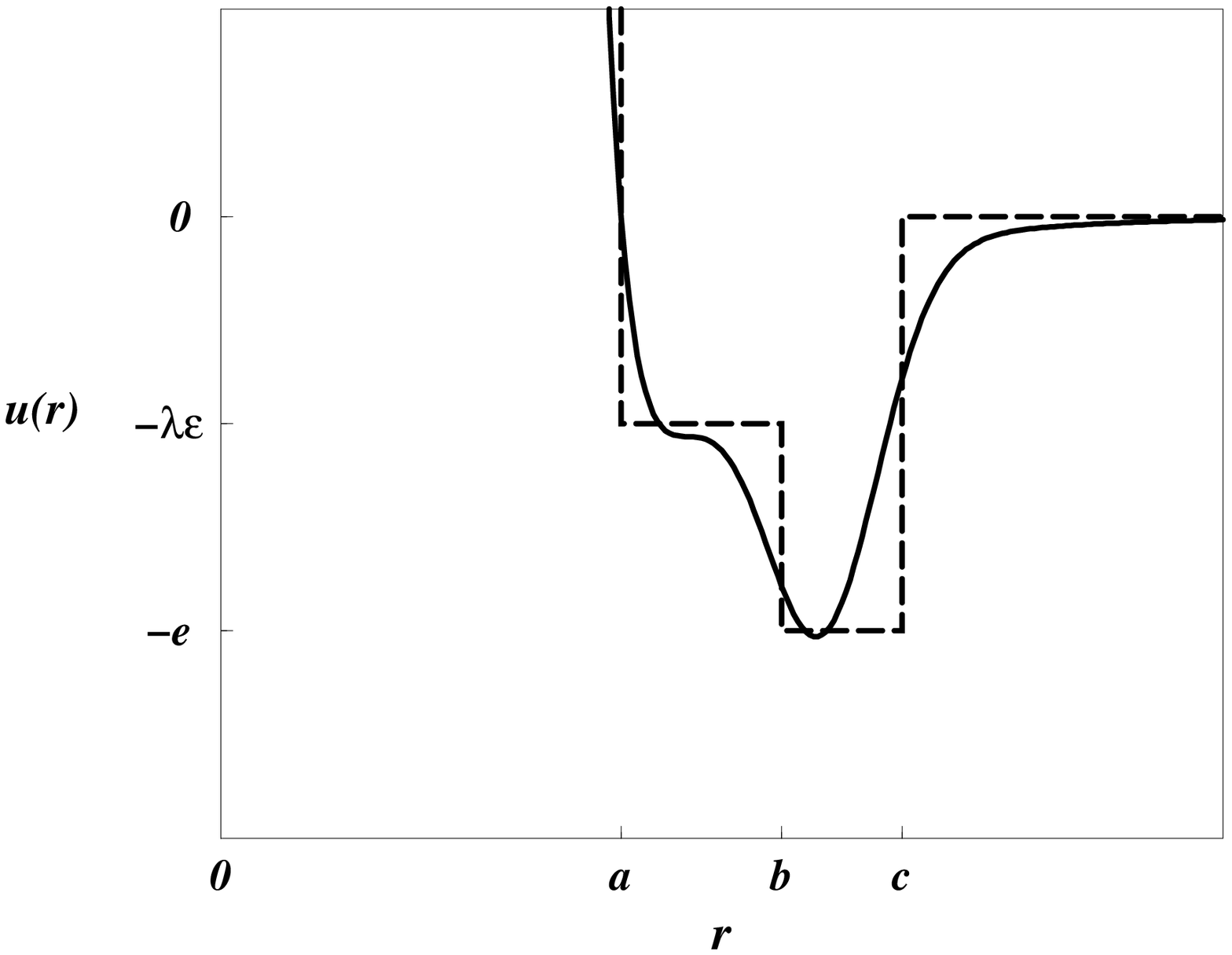}
  \hspace*{0.3cm}
  }
   }  
\caption{Discrete (broken line) and smooth (solid line) forms of the
core-softened potential $u(r)$ studied here. Length parameters $a,b,c$
and energy parameters $\epsilon, \lambda$ are shown.}
\label{figpots}
\end{figure}

In order to investigate these anomalies, we utilize computer simulation
of a class of potentials called ``core softened'' potentials, first
introduced by Stell, Hemmer, and their coworkers \cite{Stell1}. We
define a core-softened potential as a spherically symmetric potential
that has a region of negative curvature in its repulsive core
\cite{Stell3}. An example of a discrete and of a smooth core-softened
potential is shown in Fig.~\ref{figpots}. Debenedetti et al. noted that
a ``softened core'' can lead to a density anomaly \cite{Debenedetti},
i.e., one of the anomalies found in water. Furthermore, {\it ab
initio\/} calculation \cite{Liquid-Metals} and inversion of the
experimental oxygen-oxygen radial distribution function reveals that a
``core-softened'' potential can be considered a realistic first-order
approximation for the interaction between water molecules
\cite{Head-Gordon}.

Although directional bonding is certainly a fundamental issue in
obtaining quantitative predictions for network-forming liquids like
water, it could be the case that core-softened potentials can be the
simplest framework to understand the physics of those anomalies. Here we
demonstrate, by means of numerical simulations for $d=2$, that the
core-softened potential can lead to anomalies in the density, in the
compressibility and in the viscosity. We also offer an explanation for
the occurrence of these three anomalies in terms of the shape of the
potential.

The paper is organized as follows: in Sec.~\ref{secpot} we define the
potentials studied and in Sec.~\ref{secmod} we describe the methods of
simulations employed. In Sec.~\ref{secden} we present the results about
the density anomaly. In Sec.~\ref{secstruc} we discuss the relation of
the structures in the solid phase and the density anomalies. In
Sec.~\ref{secdif} we present the results for the diffusion anomaly and
give an explanation of such anomaly in terms of free volume. In
Sec.~\ref{seckt} we present the results on the compressibility anomaly.
Finally, we present the overall phase diagram in Sec.~\ref{secphas} and
our conclusions and comments in Sec.~\ref{secconc}.

\section{Discrete and Smooth Models}
\label{secpot}

\subsection{Discrete Potential}

The core-softened potentials that we study are shown in Fig.~\ref{uform}
as a function of particle pair distance $r$. The discrete potential is
composed of a hard core of diameter $a$ which has a repulsive shoulder
of width $b-a$ at depth $\lambda \epsilon$, and an attractive well of
width $c-b$ and depth $\epsilon$. The form of the function is thus

\begin{equation}
\label{uform}
 u(r)=\left\{ \begin{array}{ll} \infty &0<r<a \\ -\lambda \epsilon
	&a<r<b\\ -\epsilon &b<r<c\\ 0 &r>c \end{array} \right.
\end{equation}

All of the results reported here for the discrete potential are for
$a=1$, $b=\sqrt 2$, $c=\sqrt 3$, $\epsilon=2$, and $\lambda=0.5$.

In the case of water, one can attribute the larger distance $r=b$ to
hydrogen bonding, for which the system acquires a low energy and expands
at the same time. The inner distance $r=a$ on the other hand corresponds
to a non-hydrogen bonded energy state. Recent studies have proposed this
form of potential as the interaction between clusters of strongly bonded
pentamers of water \cite{Canpolat}. This type of interaction is expected
to reproduce the density anomaly. The reason is that at low pressures
and at low temperatures, nearest neighbor pairs sit in the outer well
which has a lower energy. By increasing $T$, in order to gain more
entropy, the system explores a larger portion of the configurational
space, which is not probed at lower temperatures. This includes
penetrating the ``soft'' core ($a<r<b$), which on average can lead to
anomalous contraction upon heating.

\subsection{Smooth Potential}

While the discrete potential $u(r)$ is appropriate for deriving the
closed form of equation of state in $1d$
\cite{Yoshimura,SadrPRL,Sadr1d}, for simulations it is not necessarily
the most appropriate. As we will show in the next section, the smooth
version of the potential $u'(r)$ requires a different method of
simulation from that of $u(r)$. The potential $u'(r)$ we use is obtained
by adding a Gaussian well to the Lennard-Jones potential and has the
form
\begin{equation}
\label{uformcont}
u'(r)=4\epsilon'\left[\left({\sigma\over r}\right)r^{12}-\left({\sigma\over
r}\right)r^{6}\right]-\lambda'\epsilon'\exp\left[-w^n\left({r\over\sigma}-{r_0\over
\sigma}\right)^n\right],
\end{equation}
for $r\leq r_c$ and vanishes for $r>r_c$.  We use $\epsilon'=1.0$,
$\lambda'=1.7$, $w=5.0$, $r_0=1.5$, $\sigma=1$, $n=2$ in order to mimic
the shape of the discrete potential, as shown in Fig.~\ref{figpots}.

\section{Molecular Dynamics Simulation Methods}
\label{secmod}

The method of simulation in both the discrete and the smooth cases is
the molecular dynamics (MD) method. Our simulations are performed in
$2d$ with periodic boundary conditions. The overall qualitative results
of the simulations for the discrete and the smooth potential are
similar, while the quantitative results differ. In what follows, we
explain in more detail the MD method used in each case.

\subsection{Constant-Volume Simulation of the Discrete Potential}
\label{ssecicdis}

For the discrete potential [Eq.~(\ref{uform})], we use the collision
table technique \cite{Allen-Tildesley} for $N=896$ disks. To each disk
we assign a radius $a/2$. We define the density $\rho$ to be the ratio
of the total area of all the disks to the area of the box. Energies are
measured in units of $\epsilon$, temperature is calculated in units of
energy divided by the Boltzmann constant, $\epsilon/k_{B}$, and the mass
of the particle is set at $m=1$.

The average pressure is calculated using the virial equation 
for step potentials~\cite{Allen-Tildesley}
\begin{equation}
 P = \left\langle{1\over V}
	\left[ K + \frac{1}{2 \delta t}
	\sum_{i,j}{}' m ({\vec v'}_i-{\vec v}_i) 
	({\vec r}_i-{\vec r}_j) 
	\right]
\right\rangle,
\end{equation}
where $K=\sum_{i} m{\vec v}_i^2/2$ is the total kinetic energy, $N$ is
the number of particles, $\sum'_{i,j}$ is the sum over the particle
pairs $(i,j)$ undergoing a collision in the time interval $\delta t$,
${\vec v}_i$ and ${\vec v'}_i$ are the velocities of the particle $i$
before and after a collision, ${\vec r}_i$ and ${\vec r}_j$ are the
positions of the particles $i$ and $j$ undergoing a collision at the
start of $\delta t$.
 
We simulate state points along constant-volume paths. For thermalization
we use the Berendsen method of rescaling the kinetic energy
\cite{Allen-Tildesley}. We thermalize the system for $10^5$ time units,
which corresponds to $\sim 10^6$ collisions per particle, and then
acquire data for $10^6$ time units corresponding to $\sim 10^7$
collisions per particle.

\subsection{Constant-Volume Simulation of the Smooth Potential}
\label{ssecicsmthV}

For the smooth version of the potential [Eq.~(\ref{uformcont})], we use
the velocity Verlet integrator method \cite{Allen-Tildesley} for a
system of $N=2500$ discs. We record the results in reduced units in
which $\sigma$, $\epsilon'$, $m$, and $k_B$ are all unity. We choose
$r_c=2.5$, and the length of each MD time step $\delta t=0.01$. We
assign to each particle a radius $2^{1/6}$, which corresponds to the
minimum of the Lennard-Jones potential $\sigma$, and define the density
$\rho$ to be the ratio of the total area of all the disks to the area of
the box.

In order to achieve a preset temperature we use the Berendsen method of
rescaling the velocities \cite{Allen-Tildesley}, resulting in the time
dependence
\begin{equation}
\label{T(t)}
T(t)=T_\infty+(T(0)-T_\infty)e^{-t/\tau}.
\end{equation}
where $\tau$ is a preset time constant \cite{Allen-Tildesley}. Typical
values of $\tau$ are around $10^4\delta t$.

We first thermalize the system for a time $\Delta t\approx 10 \tau$, and
we ensure that equilibrium is attained by monitoring the time dependence
of observables like $T$, $P$ and the potential energy $U$. Then we
acquire data, running the system for an additional period of time at
constant NVE conditions (micro-canonical ensemble). We calculate $P$ and
$T$ and we consider the system to be in equilibrium only when the
fluctuations of these quantities are less than $1 \%$ of their average
values. The acquisition time is chosen to be more than the time it takes
for the system to equilibrate and is typically $(5\times 10^4)\delta t$
to $(2\times 10^5)\delta t$.

The average simulation speed on Boston University's SGI Origin~2000
supercomputer was approximately $10\mu s$ per particle update. Each of
the state points we study requires between $8$ and $16$ hours on one
processor, and thus over $1000$ hours total computational time was
utilized.

\subsection{Constant-Pressure Simulation of the Smooth Potential}
\label{ssecibsmthP}

In order to check that our results are not biased by problems like phase
coexistence that can affect constant-volume simulations, we also perform
constant-pressure simulations in the case of the smooth potential.
Constant-pressure simulations allow us to determine more accurately the
locations of the freezing line and the density maximum. We use the
feedback method proposed by Broughton \cite{Broughton}, where the
dimensions of the box and the positions of the particles are rescaled to
obtain the desired pressure $P$. The amount of rescaling depends on the
difference between the present pressure $P(t)$ and the desired pressure
$P$.

Using the Broughton method and the Berendsen method, we gradually drive
the system to the desired $P$ and $T$, while simulating under
readjusting $V$ and $E$ conditions. We choose pressure and temperature
tolerances $\delta P$ and $\delta T$ less than $1\%$ of the desired $P$
and $T$. Once $P(t)$ and $T(t)$ reach values within the range
$P\pm\delta P$ and $T\pm\delta T$, we stop thermalization and
pressurization. If the system stays within these limits for an interval
of time of the same order of time needed to reach the desired $P$ and
$T$, we conclude that the system has equilibrated, turn off the
thermalization and pressurization, and start collecting data. During
this collection period, we monitor temperature and pressure to check
that their average values coincide with the desired ones within an
accuracy of $1\%$. For our results, the time $\Delta t$ needed to reach
equilibrium is usually of the order of 500,000 steps $\delta t$, so
$\Delta t\approx 5000$ in Lennard-Jones units. Data are acquired over a
period of $10 \Delta t$. We test our code by simulating a Lennard-Jones
system of $2304$ disks and comparing the results with the
previously-known phase diagram of a $2d$ Lennard-Jones system.

\section{Density Anomaly}
\label{secden}

\begin{figure}[htb]
\narrowtext 
\centerline{
\hbox {
  \vspace*{0.5cm} \epsfxsize=10.0cm
  \epsfbox{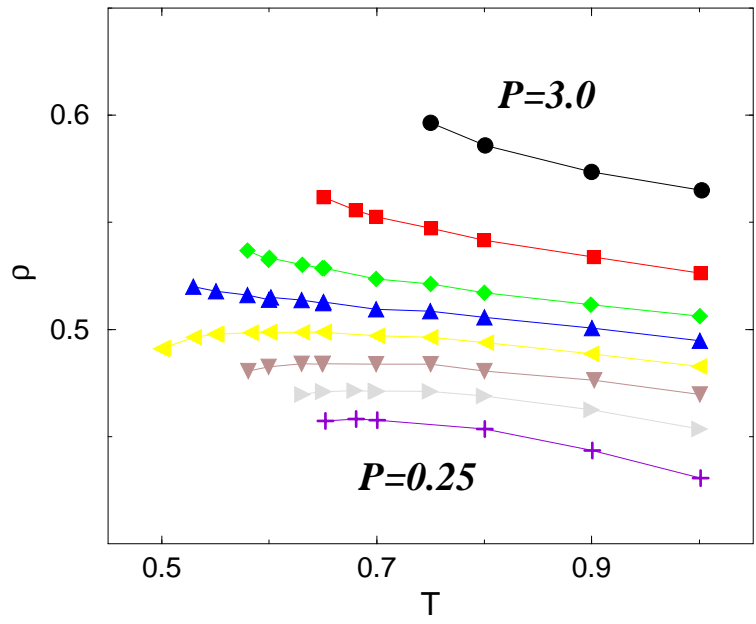} 
	\hspace*{0.3cm} 
} }
\caption{Isobars for the smooth potential}
\label{fig:isobars}
\end{figure}

The temperature of density maximum ($T_{Md}$) line is the border of the
region in the $P-T$ plane where the liquid expands upon cooling.
Fig.~\ref{fig:isobars} shows a set of different isobars for the smooth
potential. The $T_{Md}$ line corresponds to the set of maxima along
those isobars.

In the case of constant-volume simulations, the $T_{Md}$ line
corresponds to the minima along constant-volume paths ($P$ vs. $T$
graphs of Fig.~3) since for any thermodynamic quantity $X$
constant-volume paths
\begin{equation}
\label{eqsderivateives}
\left({\partial X\over\partial T}\right)_V=
\left({\partial X\over\partial T}\right)_P
-{\left({\partial X}/{\partial P}\right)_T\left({\partial V}/
{\partial T}\right)_P\over\left({\partial V}/{\partial P}\right)_T}.
\end{equation}
By substituting $X=P$ we find
\begin{equation}
\label{eqminisoch}
\left({\partial P\over\partial T}\right)_V={\alpha_P\over K_T}.
\end{equation}
where 
\begin{equation}
\label{defalphap}
\alpha_P\equiv V^{-1}\left({\partial V\over\partial T}\right)_P
\end{equation}
is the thermal expansion coefficient and
\begin{equation}
\label{defkt0}
K_T \equiv -V^{-1}\left({\partial V\over\partial P}\right)_T
\end{equation}
is the isothermal compressibility. Taking a derivative of
Eq.~(\ref{eqsderivateives}), using $\alpha_P=0$ at the $T_{md}$ line, we
find
\begin{equation}
\label{eqsecondderiv}
\left({\partial^2 P\over\partial T^2}\right)_V=
{V^{-1}\left({\partial^2 V}/{\partial T^2}\right)_P\over K_T}.
\end{equation}

\begin{figure}[htb]
\narrowtext
\centerline{
\hbox { 
  \vspace*{0.5cm} \epsfxsize=9.4cm
  \hspace*{0.3cm}
  \epsfbox{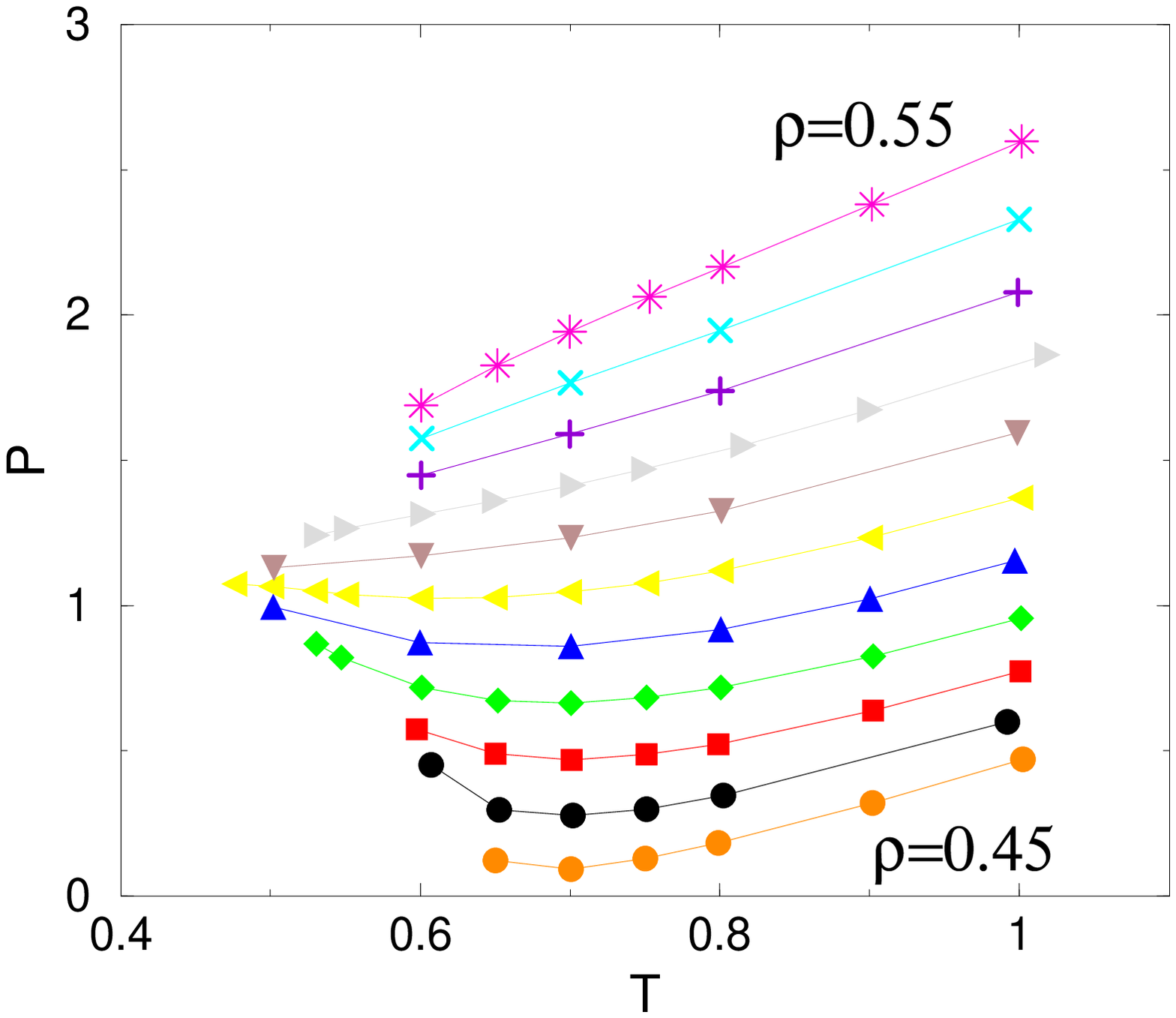} 
  }
   } 

\vspace*{0.5cm}

\centerline{
\hbox {
  \vspace*{0.5cm} \epsfxsize=10.0cm
  \epsfbox{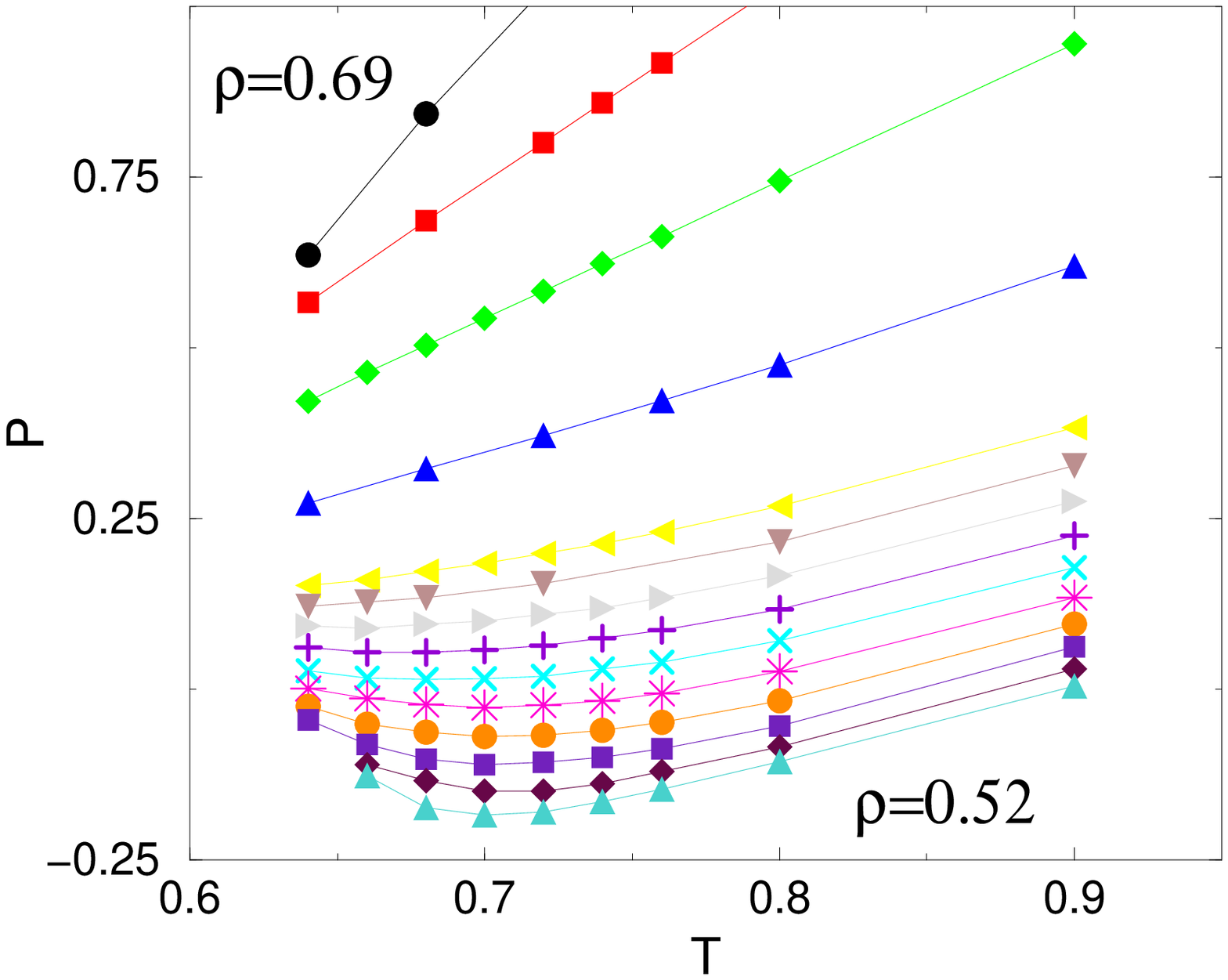} 
	}
	 }

\caption{Isochores for the smooth potential (left panel) and the
discrete potential (right panel).}
\label{fig:isochores}
\end{figure}

Equations~(\ref{eqminisoch}) and (\ref{eqsecondderiv}) show that since
$K_T$ is always positive and finite for systems in equilibrium
not at a critical point, a minimum along the isochore is equivalent
to a minimum along an isobar, which is the density maximum point
$T_{Md}$. Figure~\ref{fig:isochores} shows the isochores for the
smooth and the discrete potentials.

\begin{figure}[htb]
\narrowtext 
\centerline{
\hbox {
  \vspace*{0.5cm} \epsfxsize=11.0cm
  \epsfbox{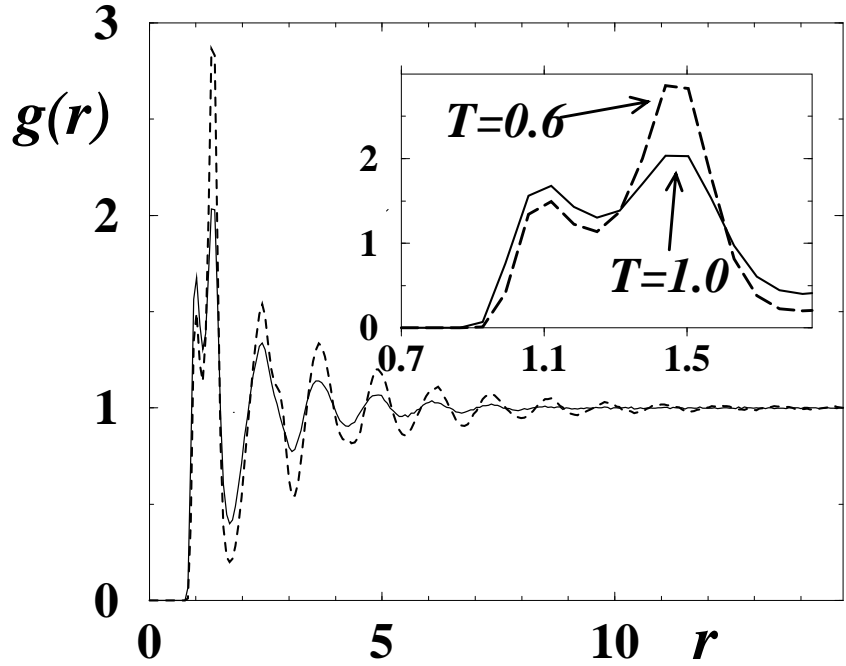} 
	\hspace*{0.3cm} 
} }

\vspace*{1cm} 
\caption{Radial distribution function at high and low temperatures,
along the P=0.48 isobar for the smooth potential. Notice how, by
lowering $T$, long range correlations develop ($g(r)=1$ if particles at
distance $r$ are uncorrelated) and more particles are expelled from
inside the soft core $r\sim 1.1$ into the attractive well $r\sim1.5$
(inset). As the average interparticle distance is growing upon cooling,
the system is expanding and there is hence a density anomaly.}
\label{fig:gidierre}
\end{figure}

To confirm that we are investigating the liquid state part of the phase
diagram, we introduce a criterion to distinguish the liquid state from a
frozen state. We determine the freezing line as the location of points
where isochores overlap. In this way we establish an approximate
location for the freezing line. Crossing this line from the liquid side,
we find a sharp decrease in $D$ coinciding with the appearance of
slowly-decaying peaks in $g(r)$ as a function of $r$, which signals the
build up of long-range correlations (Fig.~\ref{fig:gidierre}), which is
a characteristic of 2D solids.

We confirm the above criterion adopted to locate the freezing lines by
using isobaric simulations for the smooth potential. Indeed, they show a
sharp change of density, in correspondence with the estimated freezing
line at high pressures~(Fig.~\ref{fig:rhojump}, lower panel). The
presence of a hysteresis loop~(Fig.~\ref{fig:presloop}) suggests that
the liquid-solid transition is first-order; however, by lowering the
pressure the loops become less and less pronounced and eventually
disappear, leaving the possibility of an hexatic second-order transition
\cite{Nelson}.

For a few state points near the freezing line, we have checked our
results by simulating $N=2500$ particles in rectangular boxes with
aspect ratio $\sqrt 3/2$ $(L_x = \sqrt 3/2 L_y$, with $L_x \times L_y
\equiv L_0^2$) which accommodate triangular lattices perfectly. This
eliminates any possible artificial hindrance in crystallization that may
arise from the asymmetry imposed by the shape of the square box.

For water, the locus of the $T_{Md}$ line in the P-T phase diagram is of
special interest to distinguish between different scenarios proposed to
explain its anomalies \cite{dieciocho,diecinueve,veinte}. In
Fig.~\ref{phdiag} we see that the $T_{Md}$ line changes from negative
slope at high pressures to positive slope at low pressures. This change
in slope is similar to what is observed in simulating model potentials
of water like SPC/E or ST2 \cite{dieciocho,39}. These results suggest
that the change in slope can be a general phenomenon stemming from the
general core-softened form of the interaction in the simulation.

\section{Structures in the Liquid and Solid Phases}
\label{secstruc}

\begin{figure}[htb]
\narrowtext
\centerline{
\hbox { 
  \vspace*{0.5cm} \epsfxsize=10.0cm
  \hspace*{0.25cm}
  \epsfbox{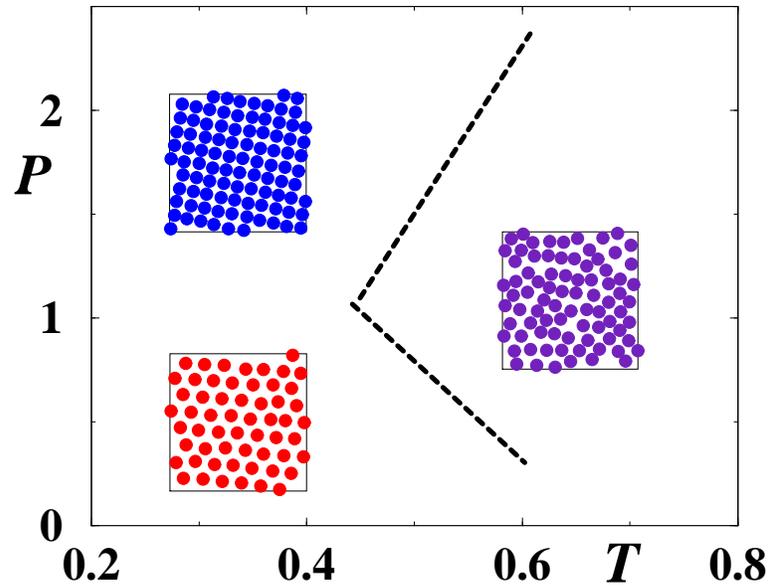} 
  }
   } 

\vspace*{0.5cm}

\centerline{
\hbox {
  \vspace*{0.5cm} \epsfxsize=11.5cm
  \epsfbox{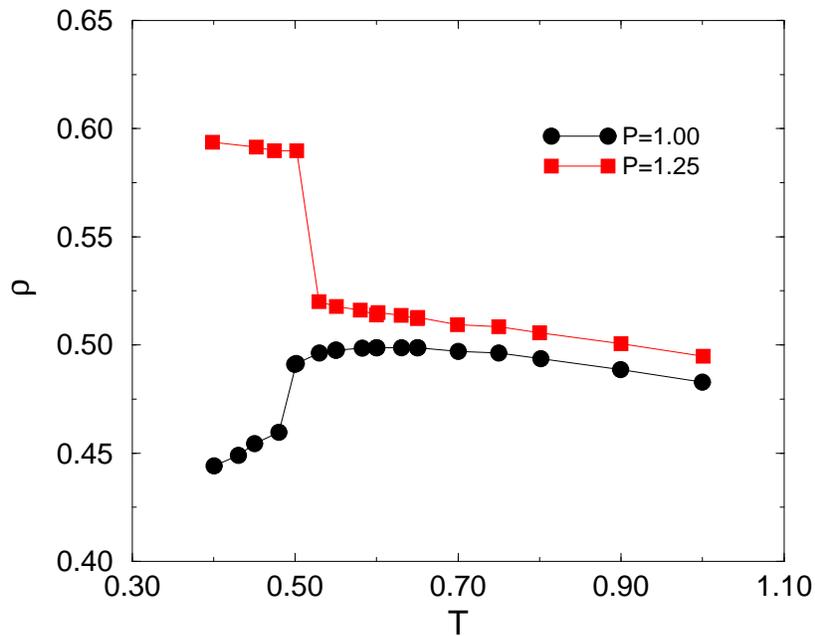} 
	}
	 }

\caption{Phases for the core-softened model (smooth potential). The
upper panel shows a snapshot of the liquid phase and snapshots of
different types of crystals for the solid phase at low pressures where
the freezing line is negatively sloped, and at high pressures where the
freezing line is positively sloped. Lower panel shows the density jumps
along isobars; note that the low pressure isobar shows a density anomaly
before jumping to a low density solid.}
\label{fig:rhojump}
\end{figure}

\begin{figure}[htb]
\narrowtext 
\centerline{
\hbox {
  \vspace*{0.5cm} \epsfxsize=11.0cm
  \epsfbox{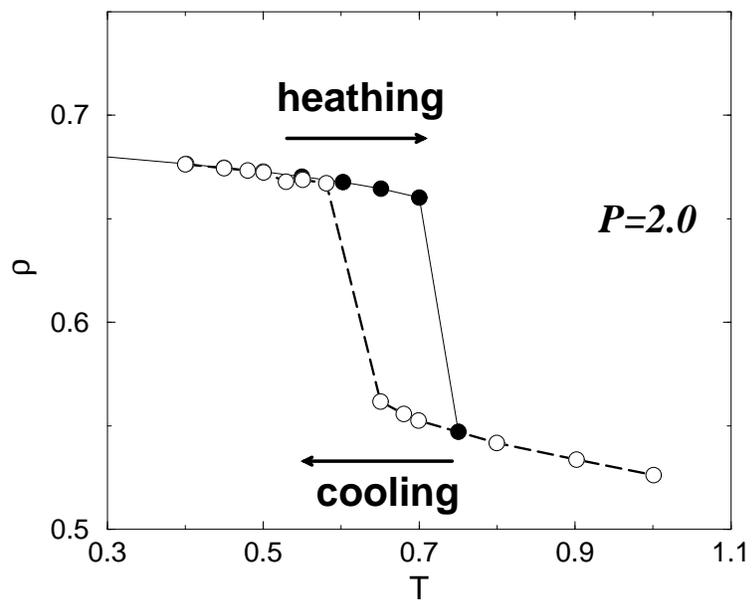} 
	\hspace*{0.3cm} 
} }
\caption{Hysteresis loop near the freezing line for a high pressure
isobar (smooth potential). The continuous line is obtained upon heating,
the dashed line upon cooling.}
\label{fig:presloop}
\end{figure}

Figure~\ref{fig:rhojump} (upper panel) shows the different phases of the
system. In our simulations we see that the $T_{Md}$ line is located in
the region of pressures where the freezing line is negatively sloped,
 as in water. A density anomaly and a negatively-sloped melting line are often
associated \cite{Debenedetti-book,StellBirthday}. This has proven to be
the case for substances like water (Fig.~\ref{waterdiff}) and
tellurium \cite{Yoshimura} and for computer models \cite{SadrPRL,Jagla}.
This association is plausible since the isobaric thermal expansion
coefficient $\alpha_P$ is related to the cross fluctuations in volume
and entropy as
\begin{equation}
\label{crossalpha}
\alpha_P \equiv \beta\langle\delta V\delta S\rangle.
\end{equation}

Approaching a freezing line, we expect local density fluctuations to
have structures similar to the neighboring solid as they are going to
trigger the liquid-solid transition. On the other hand, the
Clausius-Clapeyron relation for the slope of the freezing line
\begin{equation}
\label{ClausClap}
{dP\over dT} = {\Delta S\over\Delta V}
\end{equation}
implies that, if the freezing line is negatively sloped, the solid,
which has a lower entropy than the liquid, will have a higher specific
volume. Therefore, if the fluctuations in the liquid are ``solid-like,''
$\alpha_P$ [Eq.~(\ref{crossalpha})] will turn out to be negative.

To distinguish different local structures in the liquid, we plot the
radial distribution function $g(r)$ for different pressures and
temperatures (Fig.~\ref{fig:gidierre}). As expected, at low pressures
cooling expels particles from the core, while increasing pressure at
fixed temperature has the opposite effect.

Since our system is two-dimensional, we can use visual inspection to
develop an intuitive picture of the possible local structures
(Fig.~\ref{fig:rhojump}, upper panel). If the fluctuations in the liquid
are ``solid-like,'' near the freezing line we expect to see local
structures that resemble the structure of the nearby solid.

We find that at low $P$ and $T$, the system is frozen with a hexagonal
structure (Fig.~\ref{seelocalstruct}, lower left panel). A ``snapshot''
of the system along the same isobaric line (Fig.~\ref{seelocalstruct},
lower right panel) shows clearly that local patches with hexagonal order
are present in the liquid phase near the freezing line. We will refer to
this structure as the ``open structure.'' Similarly, at high pressures
the local patches in the liquid phase near the freezing line
(Fig.~\ref{seelocalstruct}, upper right panel) resemble the structure of
the system when it is frozen at low $T$ and high $P$
(Fig.~\ref{seelocalstruct}, upper left panel). We will refer to this as
the ``dense structure.'' For the open structures, each particle has six
neighbors sitting in the deepest well, and the softened-core behaves as
the effective core for the particles. The dense structure is the next
energetically favorable local arrangement, with four neighbors in the
external well and four in the softened core, for which the effective
core is the hard core.

\section{Diffusion Anomaly}
\label{secdif}

We next study the diffusion anomaly, which is another surprising feature
of water. While for most materials diffusivity decreases with pressure,
liquid water has an opposite behavior in a large region of the phase
diagram \cite{WaterData} (Fig.~\ref{waterdiff}). The pressures where the
system has a maximum diffusivity along isotherms define the line of the
pressure of maximum diffusivity, $P_{MD}$.

We observe that our core-softened potential reproduces this anomaly. We
first measure the mean square displacement $\langle\Delta
r^2(t)\rangle\equiv\langle[r(t+t_0)-r(t_0)]^2\rangle$ and then the
diffusion coefficient using the relation
\begin{equation}
\label{Deinst}
D=\frac{1}{2d}\lim_{t\to\infty}{\langle\Delta r^2(t)\rangle\over t}.
\end{equation} 
We measure $\langle\Delta r^2(t)\rangle$ by averaging over the starting
time $t_0$ in Eq.~(\ref{Deinst}). We find that there is a region of the
phase diagram in which $D$ increases upon increasing
$P$~(Fig~\ref{2ddiffs}).

In order to understand the diffusion anomaly we first note that, for
normal liquids, $D$ decreases with $P$ because upon increasing $P$ the
density increases and molecules are more constrained. As a result the
viscosity increases and hence $D$ decreases. In the case of water the
anomaly can be related to the fact that increasing pressure (and hence
density) breaks hydrogen bonds, which in turn increases the mobility of
the molecules. We present a more general explanation which can equally
apply to our radially symmetric core-softened interaction which does not
possess any directional bonds similar to hydrogen bonds. The low energy
inter-particle state at $r\approx b$ plays the role of non-directional
bond. Note that $D$ is proportional to the mean free path of particles,
which increases with the free volume per particle $v_{\mbox{\scriptsize
free}}\equiv v-v_{\mbox{\scriptsize ex}}$, where $v_{\mbox{\scriptsize
ex}}$ is the excluded volume per particle resulting from the effective
hard core. At low temperatures, $v_{\mbox{\scriptsize ex}}$ for the
dense structure is proportional to the area $a^2$ of the hard core,
while for the open structure it is proportional to the area $b^2$ of the
soft core. Increasing $P$ decreases $v$, which is the main effect in
normal liquids. For the core-softened liquid, on the other hand,
increasing $P$ can also decrease $v_{\mbox{\scriptsize ex}}$ by
transforming some of local open structures to dense structures. Since
both $\Delta v$ and $\Delta v_{\mbox{\scriptsize ex}}$ decrease with $P$
and since $\Delta v_{\mbox{\scriptsize free}}=\Delta v - \Delta
v_{\mbox{\scriptsize ex}}$, the effect of $P$ on $D$ depends on whether
$\Delta v$ or $\Delta v_{\mbox{\scriptsize ex}}$ dominates. The
anomalous increase in $D$ along the isotherms near the freezing line is
a sign of the dominance of the $\Delta v_{\mbox{\scriptsize ex}}$
term. Thus the anomaly in $D$ must disappear near a certain pressure
above which the average distance between particles corresponds to the
dense structure, and as a result the contribution of the open structure
to $v_{\mbox{\scriptsize ex}}$ is negligible.

We verified this in our simulation by observing a correspondence between
the disappearance of the diffusion anomaly and the disappearance of the
peak in $g(r)$, corresponding to the open structure that is observed in
real water.

\section{Isothermal Compressibility}
\label{seckt}

In order to investigate the anomaly in isothermal compressibility $K_T$,
we calculate $K_T$ at each state point using the data in Fig.~3. In the
smooth potential case we verify these results using
\begin{equation}
\label{ktsq}
K_T=\lim_{q\to0}\frac{S(q)}{n k_BT}
\end{equation}
as an alternative method \cite{chaikin-book}, where $n$ is the density
of the system and $S(q)$ is the structure function and is related to the
pair correlation function via
\begin{equation}
\label{sqg}
S(q)=1+n\int e^{i\mbox{\bf q.x}} g(\mbox{\bf x}) d\mbox{\bf x}.
\end{equation}.

We first calculate $g(r)$ for each state point averaging over all
thermalized configurations. We then perform numerical integration using
Eq.~(\ref{sqg}) to find $S(q)$, and finally we extrapolate $S(q)$ to
$q=0$ and substitute in Eq.~(\ref{ktsq}) to find $K_T$. We show an
example of the graphs for $g(r)$ and the resulting $S(q)$ (normalized by
the extra factors in Eq.~\ref{ktsq}) in Fig.~\ref{fig:gr-and-sq}. We fit
the low $q$ tail of the curve to a line to find the limiting value in
Eq.~(\ref{ktsq}).

We graph $K_T$ along isobars, as shown in Fig.~\ref{2dktsmooth}. For
large $T$, the $K_T$ decreases upon increasing $P$. For small $T$ the
behavior is the opposite and the compressibility anomaly occurs. As
seen for all isobars shown in Fig.~\ref{2dktsmooth} (except a low
pressure one), $K_T$ increases by lowering $T$.

\section{Phase Diagram}
\label{secphas}

\begin{figure}[htb]
\narrowtext
\centerline{
\hbox { 
  \vspace*{0.5cm} \epsfxsize=9.8cm
  \hspace*{0.3cm}
  \epsfbox{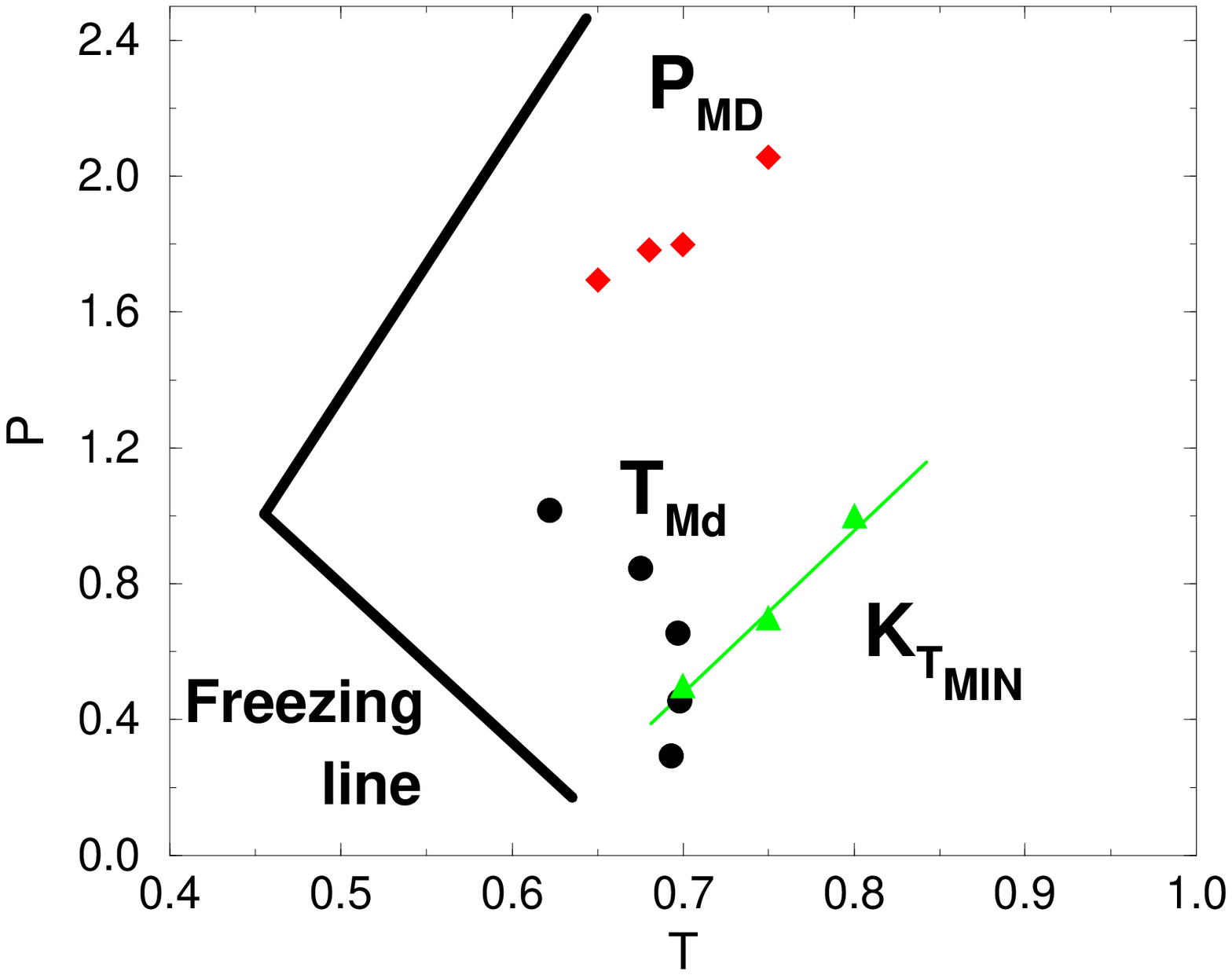} 
  }
   } 

\vspace*{0.5cm}

\centerline{
\hbox {
  \vspace*{0.5cm} \epsfxsize=10.0cm
  \epsfbox{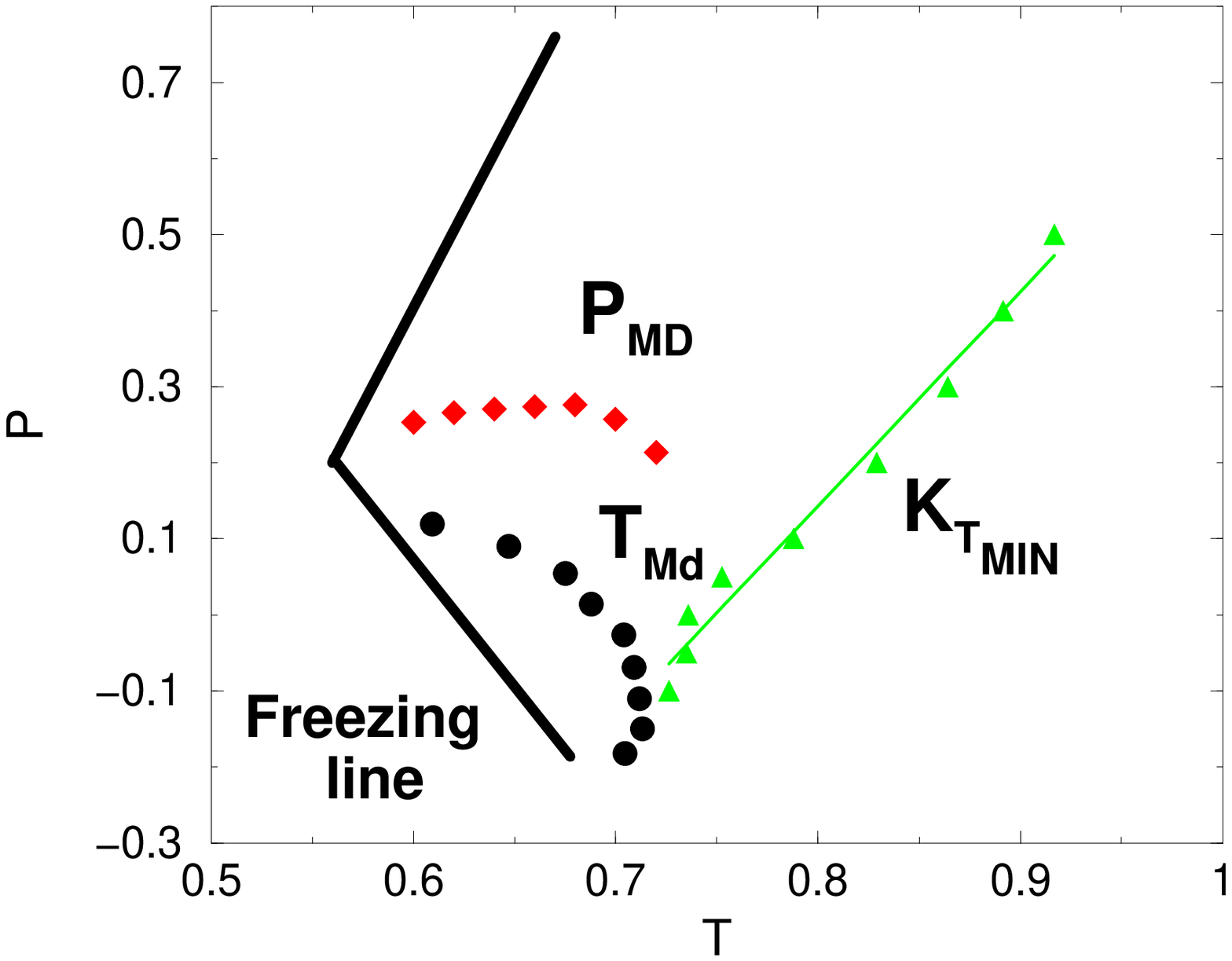} 
	}
	 }

\caption{Phase diagram with the $T_{Md}$, $P_{MD}$, $K_{T_{MIN}}$ and
freezing line for the smooth potential (left panel) and discrete
potential (right panel).}
\label{phdiag}
\end{figure}

\begin{figure}[htb]

\narrowtext \centerline{
\hbox {
  \vspace*{0.5cm}
  \epsfxsize=10.0cm
  \epsfbox{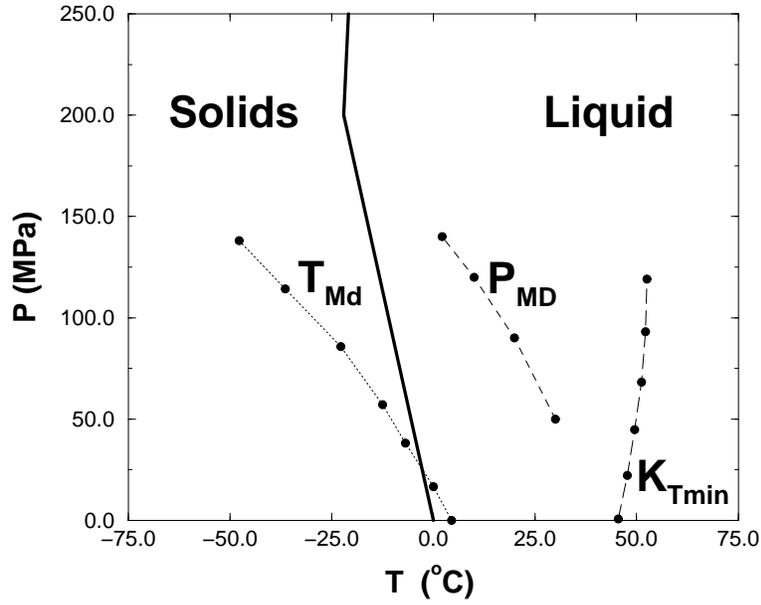}
  \hspace*{0.3cm}
  }
   }

\caption{Sketch of the phase diagram of water. The portion of the
$T_{\mbox{\scriptsize Md}}$ line that is to the left of the melting line
corresponds to experiments in the supercooled region of water. Notice
that the presence of a density anomaly in the region of the negatively
sloped melting line can occur in the metastable phase of the
liquid. Data are obtained from Refs.~\protect\cite{WaterData}.}
\label{waterdiff}
\end{figure}

\begin{figure}[htb]
\narrowtext \centerline{
\hbox {
  \vspace*{0.5cm}
  \epsfxsize=11.5cm
  \epsfbox{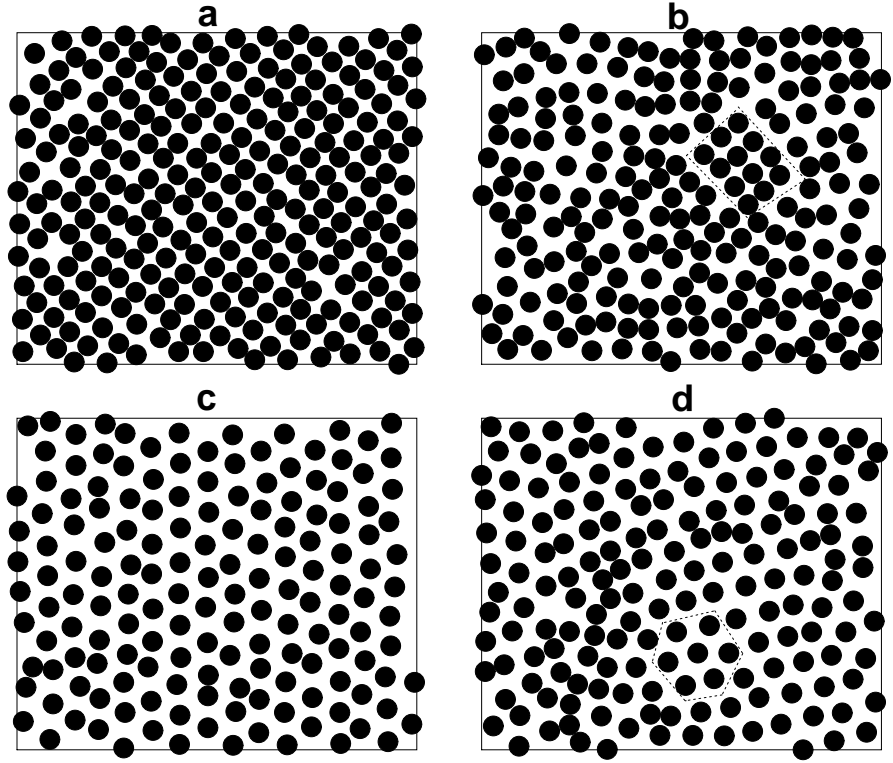}
  \hspace*{0.3cm}
  }
   }  
\vspace*{1.0cm}

\caption{ Snapshots of the system (discrete potential) in the solid phase at
high pressure (upper left) and low pressure (lower left), and in the 
liquid phase at high pressure (upper right) and low pressure (lower
right). Moving along an isobar, patches of local order
similar to the low temperature solid develop. At high pressure, the average
distance between particles (which is the radius of the disk) is of the
order of the hard core, while in the low pressure solid, the distance between
particles is larger, of the order of the softened core.}
\label{seelocalstruct}
\end{figure}

\begin{figure}[htb]
\narrowtext
\centerline{
\hbox { 
  \vspace*{0.5cm} \epsfxsize=9.8cm
  \hspace*{0.3cm}
  \epsfbox{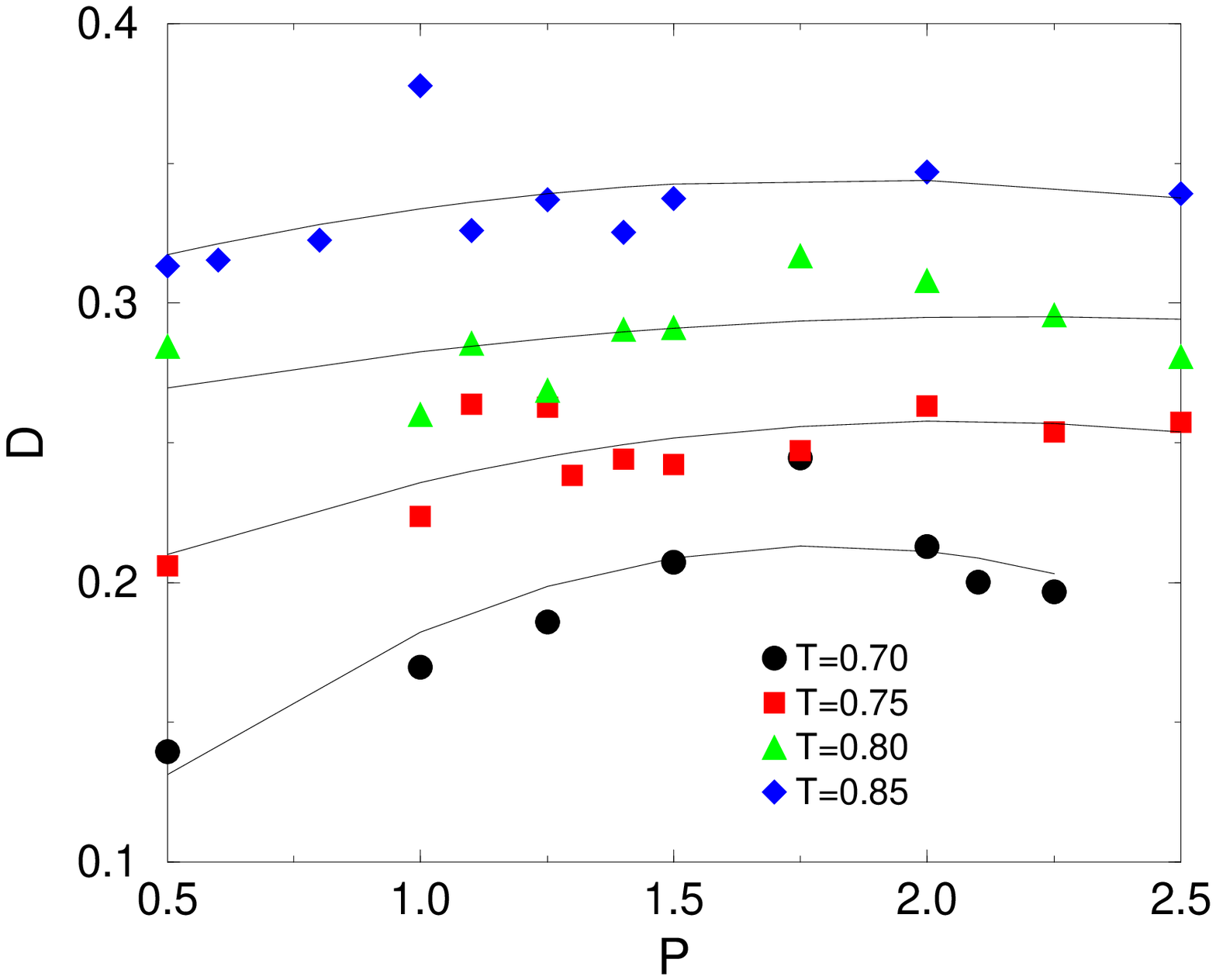} 
  }
   } 

\vspace*{0.5cm}

\centerline{
\hbox {
  \vspace*{0.5cm} \epsfxsize=10.0cm
  \epsfbox{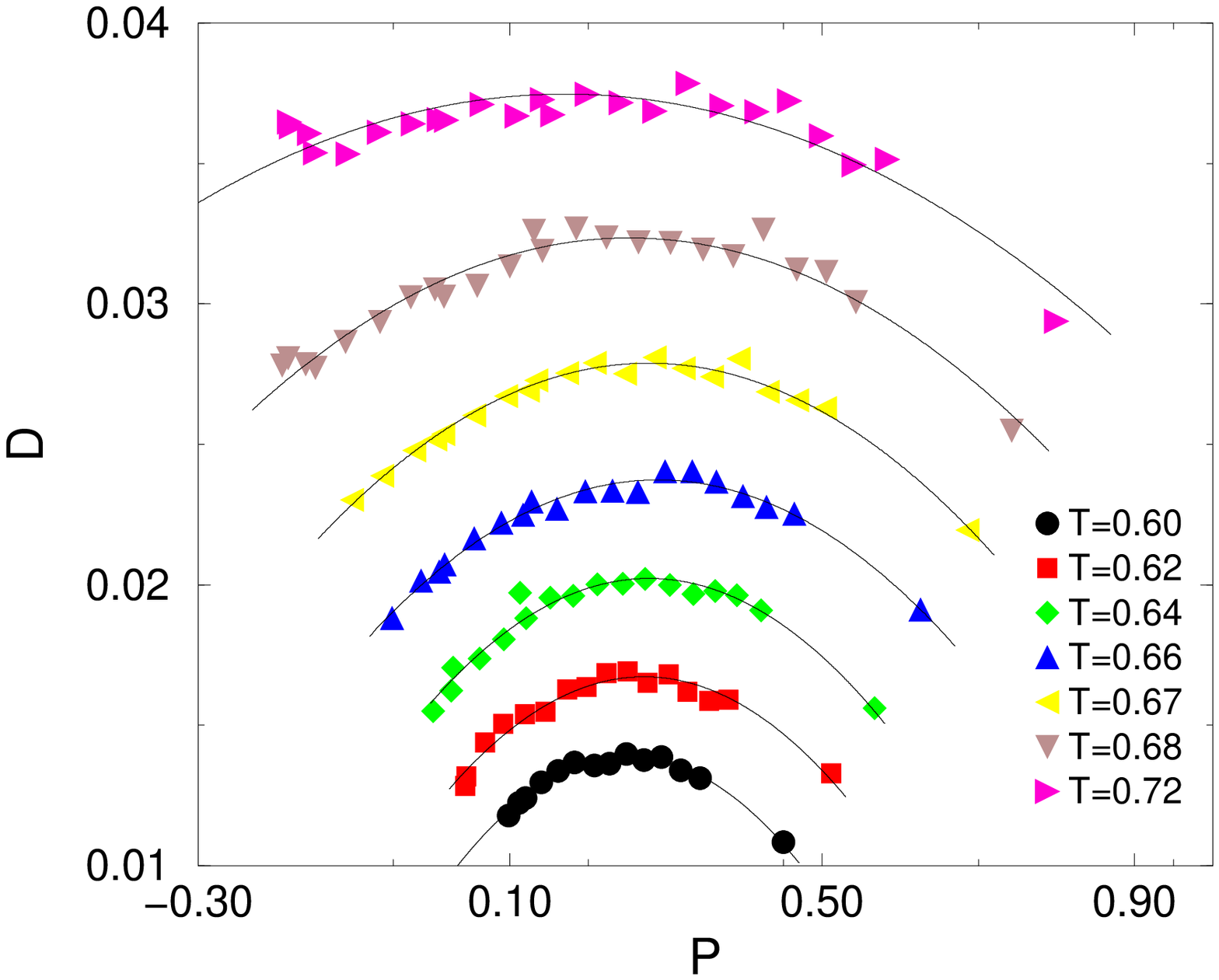} 
	}
	 }

\caption{Diffusion coefficient $D$ in the liquid phase for the
continuous potential (left panel), the discrete potential (right panel)
along various isotherms. Lines are intended as a guide for the
eye. Notice the anomalous sections of the graph, where
$(\partial D/\partial P)_T>0$.}

\label{2ddiffs}
\end{figure}

\begin{figure}[htb]
\narrowtext
\centerline{
\hbox {\vspace*{0.5cm}
  \epsfxsize=10cm
  \epsfbox{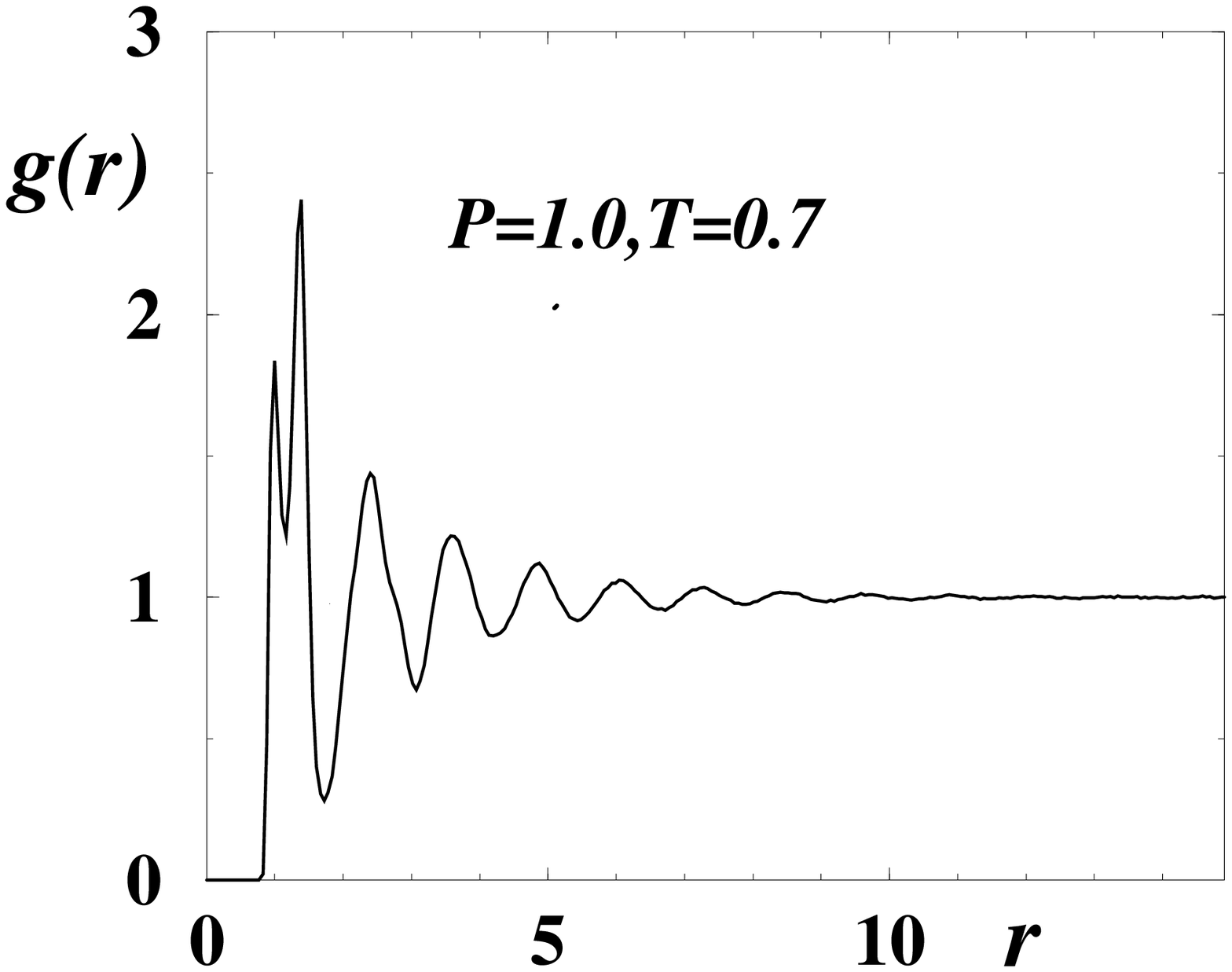}
  }
   } 

\vspace*{0.5cm}

\centerline{
\hbox {
  \epsfxsize=10cm
  \epsfbox{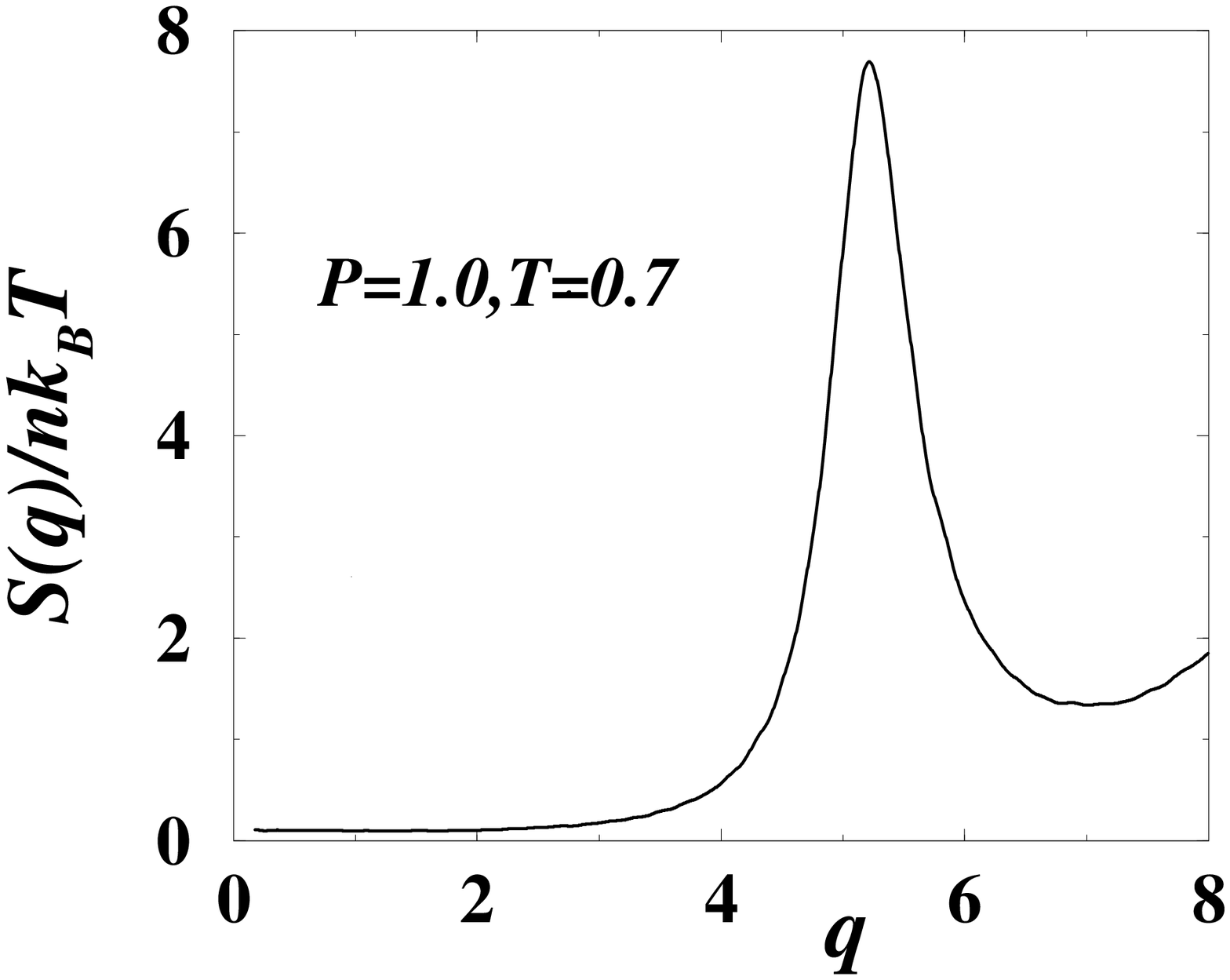}
	}
	 }

\vspace*{0.5cm}
\caption{Averaged pair distribution function for the smooth potential at
$P=1.0,T=0.7$ (upper panel). The structure function (lower panel),
multiplied by the factor $1/nk_BT$, derived by integrating $g(r)$, where
$k_B$ is the Boltzmann constant. The $q\rightarrow 0$ limit of this
function gives $K_T$, which is around $0.1$ in this case.}
\label{fig:gr-and-sq}
\end{figure}

\begin{figure}[htb]
\narrowtext
\centerline{
\hbox { 
  \vspace*{0.5cm} \epsfxsize=9.8cm
  \hspace*{0.3cm}
  \epsfbox{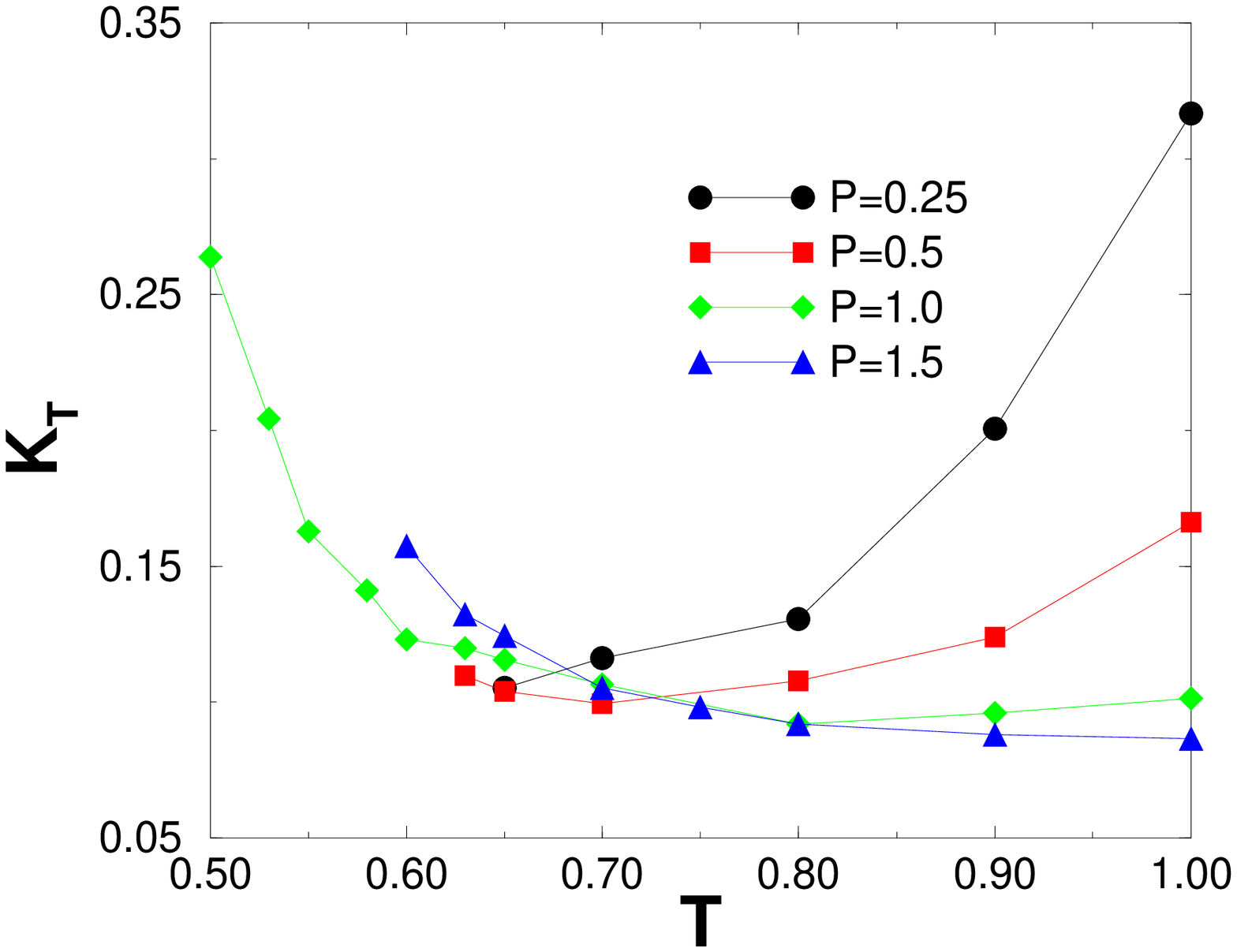} 
  }
   } 

\vspace*{0.5cm}

\centerline{
\hbox {
  \vspace*{0.5cm} \epsfxsize=10.0cm
  \epsfbox{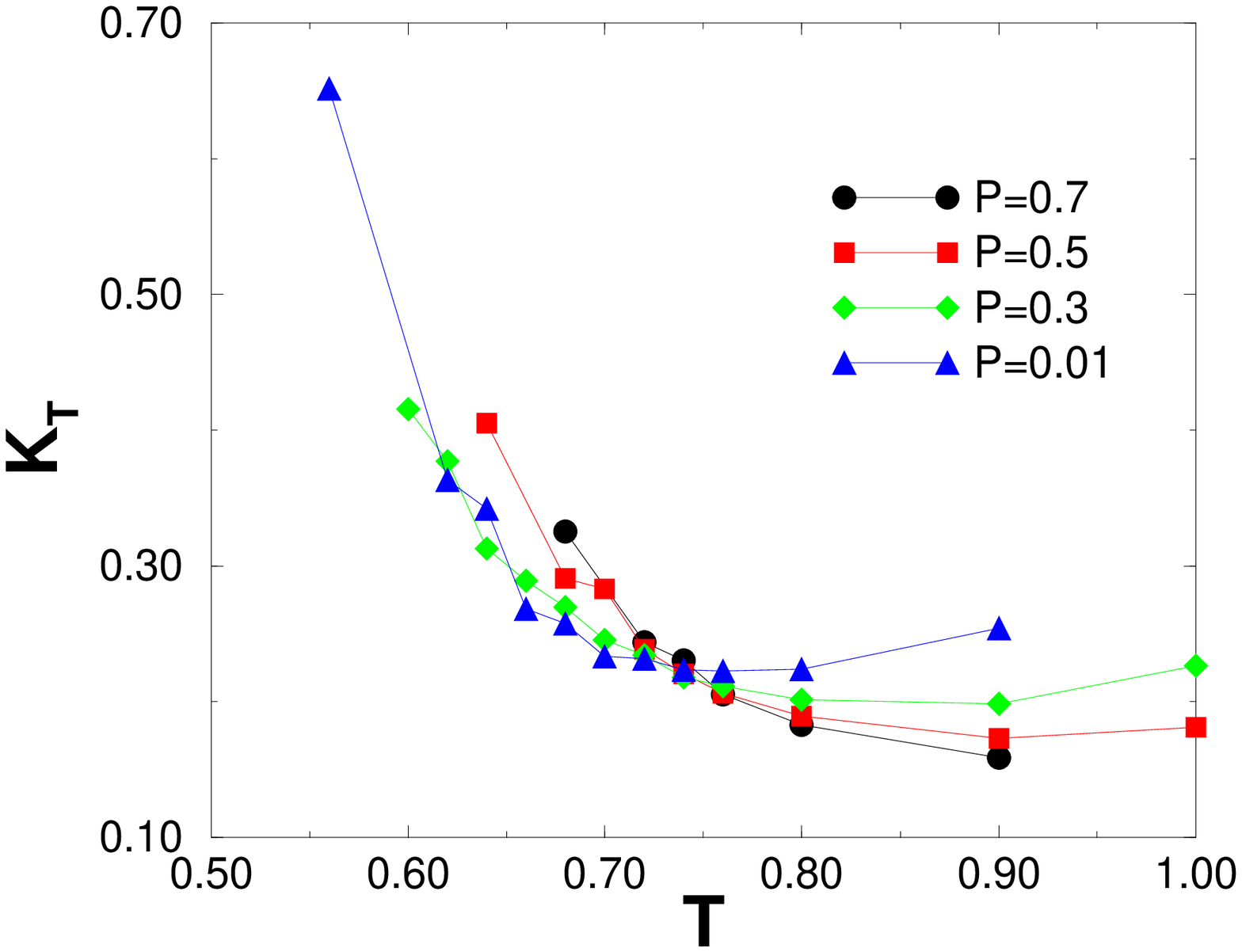} 
	}
	 }

\caption{Isothermal compressibility along isobars for the the continuous
potential (left panel) and for the discrete potential (right panel). } 
\label{2dktsmooth}
\end{figure}

In water the $T_{Md}$ line is negatively sloped for positive pressures.
For several models that mimic water behavior, it is found that the
$T_{Md}$ line has a reentrant shape, changing slope at low or negative
pressures \cite{Debenedetti-book}. In our simulations, we find such a
reentrant $T_{Md}$ line; the change of slope of the $T_{Md}$ happens at
positive pressures in the smooth version and at negative pressures in
the discrete case (Fig.~\ref{phdiag}).

Moreover, we have graphed the location of the minimum $K_T$ point along
each isobar; the locus of these points is called the minimum
compressibility line ($K_{T_{\mbox{\scriptsize min}}}$). Sastry and
coworkers \cite{diecinueve}, from basic thermodynamic arguments, show that:
(i) the $K_{T_{\mbox{\scriptsize min}}}$ line intersects the $T_{Md}$
line at its infinite slope point, and (ii) the compressibility must
increase upon cooling in the region to the left of a negatively sloped
$T_{Md}$ line. Our results are in agreement with both of these
statements~(Fig.~\ref{phdiag}).

Theories relating $D$ to the entropy \cite{AdamGibbs} predict that the
anomalous behavior $(\partial D / \partial P)_T>0$ is related to an
anomaly in the entropy~$(\partial S / \partial P)_T>0$. Due to the
Maxwell relation $(\partial S/\partial P)_T=-(\partial V/\partial T)_P$,
whenever there is a density anomaly, an entropy anomaly occurs, and the
value of entropy along isotherms reaches a maximum at the
$T_{\mbox{\scriptsize Md}}$ line.

In Fig.~\ref{phdiag} we also show the $P_{MD}$ line where $D$ reaches
its maximum with pressure. Notice that, for the continuous potential,
the maximum in $D$ shifts to higher $P$ with increasing $T$. This trend
is also observed in the $SPC/E$ model of water \cite{Sconf} but is in
contrast with the behavior of real water~(Fig.~\ref{waterdiff}).

\section{Summary}
\label{secconc}

We find that core-softened potentials reproduce the qualitative behavior
of water in many respects; in particular, the liquid phase of core-softened potentials
can show both thermodynamic anomalies and dynamic anomalies.
Moreover,
as in real water, the freezing line changes slope from a positive value
at high pressures to a negative value at low pressures in the P-T
phase diagram and more than one solid phase is present. The polymorphism
of the solid phase and the anomalies in the liquid phase can be related
to the possibility of different local structures due to the shape of the
potential.
The phase diagrams of the discrete and the smooth versions of the core
softened potential 
 are similar to that of real water,
but the $T_{Md}$ line is shifted into liquid phase and the $K_{T_{MIN}}$
line has a positive slope. However, only for the
discrete potential we find a $P_{MD}$ line with a negative slope.

\section{Acknowledgements}
\label{secack}

We thank D.~Wolf, R.~Speedy, F.~Sciortino, M.~Canpolat, F.~W.~Starr,
E.~La~Nave, M.~Meyer, A.~Skibinsky and G.~Stell for enlightening
discussions, and NSF for support.

\vspace*{1.0cm}

\newpage

\newpage

\newpage

\newpage

\newpage

\newpage

\newpage

\vspace*{1.0cm}

\newpage

\newpage

\newpage

\newpage

\end{document}